\documentstyle[prl,eqsecnum,aps]{revtex} 
\begin{document}
\author{Jian-Qi Shen\footnote{E-mail address: jqshen@coer.zju.edu.cn} and Hong-Yi Zhu}
\address{$^{a)}$State Key Laboratory of Modern Optical Instrumentation, \\
Centre for Optical and Electromagnetic Research, College of Information Science and
Engineering\\
$^{b)}$Zhejiang Institute of Modern Physics and Department of Physics \\
Zhejiang University, Spring Jade, Hangzhou 310027, People's Republic of China}
\date{\today}
\title{Exact solutions to the time-dependent supersymmetric muliphoton \\Jaynes-Cummings model
and the Chiao-Wu model\footnote{Written in May, 2002.}}
\maketitle

\begin{abstract}
By using both the Lewis-Riesenfeld invariant theory and the
invariant-related unitary transformation formulation, the present paper
obtains the exact solutions to the time-dependent supersymmetric two-level
multiphoton Jaynes-Cummings model and the Chiao-Wu model that describes the
propagation of a photon inside the optical fiber. On the basis of the fact
that the two-level multiphoton Jaynes-Cummings model possesses the
supersymmetric structure, an invariant is constructed in terms of the
supersymmetric generators by working in the sub-Hilbert-space corresponding
to a particular eigenvalue of the conserved supersymmetric generators (
{\it i.e.}, the time-independent invariant). By constructing the effective
Hamiltonian that describes the interaction of the photon with the medium of
the optical fiber, it is further verified that the particular solution to
the Schr\"{o}dinger equation is the eigenfunction of the second-quantized
momentum operator of photons field. This, therefore, means that the explicit
expression (rather than the hidden form that involves the chronological
product ) for the time-evolution operator of wave function is obtained by
means of the invariant theories.


PACC: 32.80.Wr, 33.80.Wz, 03.65.Bz
\end{abstract}


\section{INTRODUCTION}

The model that describes the interaction between a two-level atom and a
quantized single-mode electromagnetic field is termed the Jaynes-Cummings
(J-C) model\cite{Jaynes}, which can be applied to the study of many quantum
effects such as the quantum collapses and revivals of the atomic inversion,
photon antibunching, squeezing of the radiation field, inversionless light
amplification, electromagnetic induced transparency\cite
{Eberly,Alexanian,Wodkiewicz,Imamolglu}, etc. By making use of the
generalized invariant theory\cite{Gao1}, we can obtain exact solutions and
geometric phase factor of the two-level J-C model whose parameters of the
Hamiltonian are totally time-dependent. Additionally, there exists another
type of J-C model (so-called two-level multiphoton Jaynes-Cummings model)
which possesses supersymmetric Lie-algebraic structure. In this
generalization of the J-C model, the atomic transitions are mediated by $k$
photons\cite{Sukumar,Kien}. Singh has shown that this model can be used to
study the multiple atom scattering of radiation and the multiphoton
emission, absorption, and the laser processes\cite{Singh}. Some authors
introduced a supersymmetric unitary transformation to diagonalize the
Hamiltonian of this J-C model and obtain the eigenfunctions of the
stationary Schr\"{o}dinger equation$\cite{Lu1,Lu2}$. In the present paper,
we generalize this method and obtain the exact solutions and the expression
for the geometric phase factor of the totally time-dependent two-level
multiphoton Jaynes-Cummings (TLMJ-C) model through the invariant-related
unitary transformation formulation.

The exact solutions and the geometric phase factor of the time-dependent
spin model have been extensively investigated by many authors. Bouchiat and
Gibbons discussed the geometric phase for the spin-$1$ system\cite{Bouchiat}%
. Datta {\sl et al.} found the exact solution for the spin- $\frac{1}{2}$
system \cite{Datta} by means of the classical Lewis-Riesenfeld theory and
Mizrahi calculated A-A phase for the spin- $\frac{1}{2}$ system\cite{Mizrahi}
in a time-dependent magnetic field. The more systematic approach to
obtaining the formally exact solution for the spin- $j$ system was proposed
by Gao {\sl et al.}\cite{Gao0} who made use of the Lewis-Riesenfeld quantum
theory \cite{Lewis}. In this work, the generators of the Hamiltonian form
the $SU(2)$ algebra. The dynamical algebraic structure of a type of general
case of spin model such as the time-dependent $L-S$ coupled system has been
investigated and a set of $SU(N)$ generators was constructed to linearize
the Hamiltonian by Cen {\sl et al.}\cite{Cen}. Investigation in the
direction of many-spin system has been completed by F. Yan et al.\cite{Yan}
who used a unitary transformation formulation which originated in the work
of Gao in 1991 \cite{Gao1}. They discussed the time evolution of the
Heisenberg spin system and obtained the formally exact solutions of the
Schr\"{o}dinger equation in a time-dependent magnetic field. The
relationship between the unitary transformations of the dynamics of quantum
systems with time-dependent Hamiltonians and the gauge theories has been
found by Montesinos {\sl et al.}\cite{Montesinos}. In particular, they
showed that the nonrelativistic dynamics of spin- $\frac{1}{2}$ particles in
a magnetic field $B^{i}(t)$ can be formulated in a natural way as $SU(2)$
gauge theory, with the magnetic field $B^{i}(t)$ playing the role of the
gauge potential. This geometric interpretation provides a powerful method to
find the exact solutions of the Schr\"{o}dinger equation. Although many
investigations have been done in the direction of the time-dependent spin
model, the subjects of all these works were the first-quantized cases. In
this paper, we consider another time-dependent second-quantized spin model,
namely, the generalized Chiao-Wu model\cite{Chiao}, which describes the
propagation of the photon inside the optical fiber.

In 1984, Berry showed that the wavefunction in quantum adiabatic process
would give rise to the geometric phase factor\cite{Berry}. Afterwards, Chiao
and Wu suggested their model, which is the first physical realization of
Berry$^{,}$s geometric phase. It is well known that the geometric phase
factor appears only in the systems with the time-dependent Hamiltonian or
the Hamiltonian involving the evolution parameters. Both the classical
geometric phase factor and the quantum geometric phase factor that is
associated with vacuum, exist in the systems of the second-quantized field
theory. In 1980$^{,}$s, there was an argument concerning whether the
geometric phase in the fiber experiment performed by Tomita and Chiao\cite
{Tomita} belongs to the quantum or classical level\cite
{Kwiat,Haldane1,Robinson,Haldane2,ibid}. Chiao-Wu$^{,}$s theory \cite{Chiao}
which is concerned with the polarization of the light propagating inside the
noncoplanar optical fiber has no expressions for the Hamiltonian; moreover,
their theory is first-quantization. From the point of view of us, it is
consequently not applicable to investigating the geometric phase factor of
quantum level. Only the system, of which whose Hamiltonian is
second-quantized will present the geometric phase factor of quantum level.
In what follows we show that the propagation of the photon in the fiber is
essentially a second-quantized problem, {\it i.e.}, the quantum-theory problem
whose Hamiltonian is time-dependent.

Note, however, that Berry$^{,}$s theory of the geometric phase proposed in
1984 is applicable only to the case of adiabatic approximation \cite{Berry}.
The invariant theory that is appropriate to treat the time-dependent systems
was first proposed by Lewis and Riesenfeld (L-R)\cite{Lewis} in 1969. In
1991, Gao {\sl et al.} generalized the L-R invariant theory and proposed the
invariant-related unitary transformation formulation\cite{Gao1,Gao3}. One of
the advantages of this unitary transformation method is that it can
transform the hidden form (which is related to the chronological product) of
the time-evolution operator $U(t)$ into the explicit expression. Many works
have shown that the invariant-related unitary transformation approach is a
powerful tool for investigating time-dependent systems and geometric phase
factor \cite{Fu2,Shen,Shen2}.

\section{THE INVARIANT THEORY AND THE INVARIANT-RELATED UNITARY
TRANSFORMATION FORMULATION}

For the sake of illustrating the L-R invariant theory\cite{Lewis} easily, we
consider a one-dimensional system whose Hamiltonian $H(t)$ is
time-dependent. According to the L-R invariant theory, a Hermitian operator $%
I(t)$ is called invariant if it satisfies the following invariant equation (
in the unit $\hbar =1$ )

\begin{equation}
\frac{\partial I(t)}{\partial t}+\frac{1}{i}[I(t),H(t)]=0.  \label{eq21}
\end{equation}
The eigenvalue equation of the time-dependent invariant $\left| \lambda
_{n},t\right\rangle $ is given
\begin{equation}
I(t)\left| \lambda _{n},t\right\rangle =\lambda _{n}\left| \lambda
_{n},t\right\rangle,  \label{eq22}
\end{equation}
where
\begin{equation}
\frac{\partial \lambda _{n}}{\partial t}=0.  \label{eq23}
\end{equation}
The time-dependent Schr\"{o}dinger equation (in the unit $\hbar =1$) for the
system is

\begin{equation}
i\frac{\partial \left| \Psi (t)\right\rangle _{s}}{\partial t}=H(t)\left|
\Psi (t)\right\rangle _{s}.  \label{eq24}
\end{equation}
In terms of the L-R invariant theory, the particular solution $\left|
\lambda _{n},t\right\rangle _{s}$ of Eq. (\ref{eq24}) differs from the
eigenfunction $\left| \lambda _{n},t\right\rangle $ of the invariant $I(t)$
only by a phase factor $\exp [\frac{1}{i}\phi _{n}(t)]$, then the general
solution of the Schr\"{o}dinger equation (\ref{eq24}) can be written as

\begin{equation}
\left| \Psi (t)\right\rangle _{s}=%
%TCIMACRO{\tsum}%
%BeginExpansion
\mathop{\textstyle\sum}%
%EndExpansion
_{n}C_{n}\exp [\frac{1}{i}\phi _{n}(t)]\left| \lambda _{n},t\right\rangle ,
\label{eq25}
\end{equation}
where

\[
\phi _{n}(t)=\int_{0}^{t}\left\langle \lambda _{n},t^{^{\prime
}}\right| H(t^{^{\prime }})-i\frac{\partial }{\partial t^{^{\prime
}}}\left| \lambda _{n},t^{^{\prime }}\right\rangle {\rm
d}t^{^{\prime }},
\]

\begin{equation}
C_{n}=\langle \lambda _{n},t=0\left| \Psi (t=0)\right\rangle _{s}.
\label{eq26}
\end{equation}
$\left| \lambda _{n},t\right\rangle _{s}=\exp [\frac{1}{i}\phi
_{n}(t)]\left| \lambda _{n},t\right\rangle $ $(n=1,2,\cdots )$ are said to
form a complete set of the solutions of Eq. (\ref{eq24}). The statement
outlined above is the basic content of the L-R invariant theory.

A time-dependent unitary transformation operator can be constructed to
transform $I(t)$ into a time-independent invariant $I_{V}\equiv V^{\dagger
}(t)I(t)V(t)$ \cite{Gao1,Gao3} with
\begin{eqnarray}
I_{V}\left| \lambda _{n}\right\rangle &=&\lambda _{n}\left| \lambda
_{n}\right\rangle ,  \label{eq27} \\
\left| \lambda _{n}\right\rangle &=&V^{\dagger }(t)\left| \lambda
_{n},t\right\rangle .  \label{eq28}
\end{eqnarray}
Under the unitary transformation $V(t),$ the Hamiltonian $H(t)$ is
correspondingly changed into $H_{V}(t)$

\begin{equation}
H_{V}(t)=V^{\dagger }(t)H(t)V(t)-V^{\dagger }(t)i\frac{\partial V(t)}{%
\partial t}.  \label{eq29}
\end{equation}
In accordance with this unitary transformation method\cite{Gao1}, it is very
easy to verify that the particular solution $\left| \lambda
_{n},t\right\rangle _{s0}$ of the time-dependent Schr\"{o}dinger equation
associated with $H_{V}(t)$

\begin{equation}
i\frac{\partial \left| \lambda _{n},t\right\rangle _{s0}}{\partial t}%
=H_{V}(t)\left| \lambda _{n},t\right\rangle _{s0}  \label{eq210}
\end{equation}
is different from the eigenfunction $\left| \lambda _{n}\right\rangle $ of $%
I_{V}$ only by the same phase factor $\exp [\frac{1}{i}\phi _{n}(t)]$ as
that in Eq. (\ref{eq25}), {\it i.e.},

\begin{equation}
\left| \lambda _{n},t\right\rangle _{s0}=\exp [\frac{1}{i}\phi
_{n}(t)]\left| \lambda _{n}\right\rangle .  \label{eq211}
\end{equation}
Substitution of $\left| \lambda _{n},t\right\rangle _{s0}$ of Eq. (\ref
{eq210}) into Eq. (\ref{eq211}) yields

\begin{equation}
\dot{\phi}(t)\left| \lambda _{n}\right\rangle =H_{V}(t)\left| \lambda
_{n}\right\rangle ,  \label{eq212}
\end{equation}
which means that $H_{V}(t)$ differs from $I_{V}(t)$ only by a time-dependent
multiplying c-number factor. It can be seen from Eq. (\ref{eq212}) that the
particular solution of Eq. (\ref{eq210}) can be easily obtained by
calculating the phase from Eq. (\ref{eq212}). Thus, one is led to the
conclusion that if the $V(t),$ $I_{V},$ $H_{V}(t)$ and the eigenfunction $%
\left| \lambda _{n}\right\rangle $ of $I_{V}$ have been found, the problem
of solving the complicated time-dependent Schr\"{o}dinger equation (\ref
{eq24}) reduces to that of solving the much simplified equation (\ref{eq210}%
). This paper obtains the exact solutions of the time-dependent
Schr\"{o}dinger equation describing TLMJ-C model and expression for its
geometric phase factor by making use of this invariant-related unitary
transformation method.

\section{THE EXACT SOLUTIONS OF THE TIME-DEPENDENT TLMJ-C MODEL}

The time-dependent Schr\"{o}dinger equation and the Hamiltonian of the
time-dependent TLMJ-C model under the rotating wave approximation are given
by
\[
H(t)\left| \Psi (t)\right\rangle =i\frac{\partial }{\partial t}\left| \Psi
(t)\right\rangle ,
\]

\begin{equation}
H(t)=\omega (t)a^{\dagger }a+\frac{\omega _{0}(t)}{2}\sigma
_{z}+g(t)(a^{\dagger })^{k}\sigma _{-}+g^{\ast }(t)a^{k}\sigma _{+}
\label{eq31}
\end{equation}
where $a^{\dagger }$ and $a$ are the creation and annihilation operators for
the electromagnetic field, and obey the commutation relation $\left[
a,a^{\dagger }\right] =1$; $\sigma _{\pm }$ and $\sigma _{z}$ denote the
two-level atom operators which satisfy the commutation relation $\left[
\sigma _{z},\sigma _{\pm }\right] =\pm 2\sigma _{\pm }$ ; $g(t)$ and $%
g^{\ast }(t)$ are the coupling coefficients and $k$ is the photon number in
each atom transition process; $\omega _{0}(t)$ and $\omega (t)$ are
respectively the transition frequency and the mode frequency. All the
parameters in $H(t)$ are time-dependent.

The supersymmetric structure can be found in the TLMJ-C model by defining
the following supersymmetric transformation generators\cite{Lu1,Lu2}:

\begin{eqnarray}
N &=&a^{\dagger }a+\frac{k-1}{2}\sigma _{z}+\frac{1}{2}=\left(
\begin{array}{cc}
a^{\dagger }a+\frac{k}{2} & 0 \\
0 & aa^{\dagger }-\frac{k}{2}
\end{array}
\right) ,\quad N^{^{\prime }}=\left(
\begin{array}{cc}
a^{k}(a^{\dagger })^{k} & 0 \\
0 & (a^{\dagger })^{k}a^{k}
\end{array}
\right) ,  \nonumber \\
Q &=&(a^{\dagger })^{k}\sigma _{-}=\left(
\begin{array}{cc}
0 & 0 \\
(a^{\dagger })^{k} & 0
\end{array}
\right) ,\quad Q^{\dagger }=a^{k}\sigma _{+}=\left(
\begin{array}{cc}
0 & a^{k} \\
0 & 0
\end{array}
\right) .  \label{eq32}
\end{eqnarray}
It is easily verified that $(N,N^{^{\prime }},Q,Q^{\dagger })$ form
supersymmetric generators and have supersymmetric Lie algebra properties,
{\it i.e.},

\begin{eqnarray}
Q^{2} &=&(Q^{\dagger })^{2}=0,\quad \left[ Q^{\dagger },Q\right]
=N^{^{\prime }}\sigma _{z},\quad \left[ N,N^{^{\prime }}\right] =0,\quad %
\left[ N,Q\right] =Q,  \nonumber \\
\left[ N,Q^{\dagger }\right] &=&-Q^{\dagger },\quad \left\{ Q^{\dagger
},Q\right\} =N^{^{\prime }},\quad \left\{ Q,\sigma _{z}\right\} =\left\{
Q^{\dagger },\sigma _{z}\right\} =0,  \nonumber \\
\left[ Q,\sigma _{z}\right] &=&2Q,\quad \left[ Q^{\dagger },\sigma _{z}%
\right] =-2Q^{\dagger },\quad \left( Q^{\dagger }-Q\right) ^{2}=-N^{^{\prime
}}  \label{eq33}
\end{eqnarray}
where $\left\{ {}\right\} $ denotes the anticommuting bracket. By the aid of
(\ref{eq32}) and (\ref{eq33}), the Hamiltonian (\ref{eq31}) of the TLMJ-C
model can be rewritten as

\begin{equation}
H(t)=\omega (t)N+\frac{\omega (t)-\delta (t)}{2}\sigma _{z}+g(t)Q+g^{\ast
}(t)Q^{\dagger }-\frac{\omega (t)}{2}  \label{eq34}
\end{equation}
with $\delta (t)=k\omega (t)-\omega _{0}(t).$

The equation which governs the time evolution of the TLMJ-C model is the
time-dependent Schr\"{o}dinger equation (\ref{eq24}). We will show the
solvability of Eq. (\ref{eq24}) by using the generalized invariant
formulation in what follows.

According to the invariant theory, we should first construct an invariant $%
I(t).$ From the invariant equation (\ref{eq21}) one can see that
$I(t)$ is the linear combination of $N,\sigma _{z},Q$ and
$Q^{\dagger }.$ However, it should be pointed out that the
generalized invariant theory can only be applied to the study of
the system for which there exists the quasi-algebra defined in
Ref.\cite{Mizrahi}. It is easily seen from (\ref{eq33}) that there
is no such quasi-algebra for the TLMJ-C model. We generalize the
method which has been used for finding the dynamical algebra
$O(4)$ of the hydrogen atom to treat this type of time-dependent
models. In the case of hydrogen, the dynamical algebra $O(4)$ was
found by working in the sub-Hilbert-space corresponding to a
particular eigenvalue of the Hamiltonian \cite{Schiff}. In this
paper, we will show that a generalized quasi-algebra can also be
found by working in a sub-Hilbert-space corresponding to a
particular eigenvalue of $N^{^{\prime }}$ in the time-dependent
TLMJ-C model . This generalized quasi-algebra enables one to
obtain the complete set of exact
solutions for the Schr\"{o}dinger equation. It is easily verified that $%
N^{^{\prime }}$ commutes with $H(t)$ and is a time-independent invariant
according to Eq. (\ref{eq21}).

Use is made of $a^{k}(a^{\dagger })^{k}\left| m\right\rangle =\frac{(m+k)!}{%
m!}\left| m\right\rangle $ and $(a^{\dagger })^{k}a^{k}\left|
m+k\right\rangle =\frac{(m+k)!}{m!}\left| m+k\right\rangle ,$ then one can
arrive at

\begin{equation}
N^{^{\prime }}{%
%TCIMACRO{\binom{\left| m\right\rangle }{\left| m+k\right\rangle }}%
%BeginExpansion
{\left| m\right\rangle  \choose \left| m+k\right\rangle }%
%EndExpansion
}=\lambda _{m}{%
%TCIMACRO{\binom{\left| m\right\rangle }{\left| m+k\right\rangle }}%
%BeginExpansion
{\left| m\right\rangle  \choose \left| m+k\right\rangle }%
%EndExpansion
}  \label{eq35}
\end{equation}
with $\lambda _{m}=\frac{(m+k)!}{m!}.$ Thus we obtain the
supersymmetric quasi-algebra $(N,Q,Q^{\dagger },\sigma _{z})$ in
the sub-Hilbert-space corresponding to the particular eigenvalue
$\lambda _{m}$ of $N^{^{\prime }}, $ by replacing the generator
$N^{^{\prime }}$ with $\lambda _{m}$ in the commutation relations
in (\ref{eq33}), namely,

\begin{equation}
\left[ Q^{\dagger },Q\right] =\lambda _{m}\sigma _{z},\quad \left\{
Q^{\dagger },Q\right\} =\lambda _{m},\quad \left( Q^{\dagger }-Q\right)
^{2}=-\lambda _{m}.  \label{eq36}
\end{equation}

In accordance with the invariant theory, the invariant $I(t)$ is often of
the form

\begin{equation}
I(t)=c(t)Q^{\dagger }+c^{\ast }(t)Q+b(t)\sigma _{z}  \label{eq37}
\end{equation}
where $c^{\ast }(t)$ is the complex conjugation of $c(t),$ and $b(t)$ is
real. Substitution of the expressions (\ref{eq34}) and (\ref{eq37}) for $%
I(t) $ and $H(t)$ into Eq. (\ref{eq21}) leads to the following set of
auxiliary equations

\begin{eqnarray}
\dot{c}-\frac{1}{i}[c\delta +2bg] &=&0,\quad \dot{c}^{\ast }+\frac{1}{i}%
[c^{\ast }\delta +2bg^{\ast }]=0,  \nonumber \\
\stackrel{.}{b}+\frac{1}{i}\lambda _{m}(c^{\ast }g-cg^{\ast }) &=&0,
\label{eq38}
\end{eqnarray}
where dot denotes the time derivative. The three time-parameters $c,c^{\ast
} $ and $b$ in $I(t)$ are determined by these three auxiliary equations.

Using the invariant-related unitary transformation method\cite{Gao1}, we
define the unitary transformation operator as follows

\begin{equation}
V(t)=\exp [\alpha (t)Q-\alpha ^{\ast }(t)Q^{\dagger }]  \label{eq39}
\end{equation}
with $\alpha ^{\ast }(t)$ being the complex conjugation of $\alpha (t).$
With the help of the commutation relations (\ref{eq33}), it can be found
that, by the complicated and lengthy computations, if $\alpha (t)$ and $%
\alpha ^{\ast }(t)$ satisfy the following equations

\begin{equation}
\sin (4\alpha \alpha ^{\ast }\lambda _{m})^{\frac{1}{2}}=\frac{\lambda
_{m}(c\alpha ^{\ast }+c^{\ast }\alpha )}{(4\alpha \alpha ^{\ast }\lambda
_{m})^{\frac{1}{2}}},\quad \cos (4\alpha \alpha ^{\ast }\lambda _{m})^{\frac{%
1}{2}}=b,  \label{eq310}
\end{equation}
a time-independent invariant can be obtained as follows

\begin{equation}
I_{V}\equiv V^{\dagger }(t)I(t)V(t)=\sigma _{z}.  \label{eq311}
\end{equation}
From Eq. (\ref{eq310}), we substitute the time-dependent parameters $\lambda
$ and $\gamma $ for $c,c^{\ast }$ and $b$ in $I(t)$ for simplicity and
convenience, and the results are

\begin{eqnarray}
\alpha &=&\frac{\frac{\lambda }{2}\exp (i\gamma )}{\lambda _{m}^{\frac{1}{2}}%
},\quad \alpha ^{\ast }=\frac{\frac{\lambda }{2}\exp (-i\gamma )}{\lambda
_{m}^{\frac{1}{2}}},  \nonumber \\
c &=&\frac{\sin \lambda \exp (i\gamma )}{\lambda _{m}^{\frac{1}{2}}},\quad
c^{\ast }=\frac{\sin \lambda \exp (-i\gamma )}{\lambda _{m}^{\frac{1}{2}}}.
\label{eq312}
\end{eqnarray}
Thus, the invariant $I(t)$ in (\ref{eq37}) can be rewritten

\begin{equation}
I(t)=\frac{\sin \lambda }{\lambda _{m}^{\frac{1}{2}}}[\exp (i\gamma )Q+\exp
(-i\gamma )Q^{\dagger }]+\cos \lambda \sigma _{z}.  \label{eq313}
\end{equation}
In the meanwhile, under the unitary transformation (\ref{eq39}), the
Hamiltonian (\ref{eq34}) can be transformed into

\begin{eqnarray}
H_{V}(t) &\equiv &V^{\dagger }(t)H(t)V(t)-V^{\dagger }(t)i\frac{\partial }{%
\partial t}V(t)  \nonumber \\
&=&\omega N-\frac{\omega }{2}+\{\frac{\omega }{2}(1-\cos \lambda )+\frac{1}{2%
}\lambda _{m}^{\frac{1}{2}}[g\exp (-i\gamma )+g^{\ast }\exp (i\gamma )]\sin
\lambda +  \nonumber \\
&&+\frac{\omega -\delta }{2}\cos \lambda +\frac{\dot{\gamma}}{2}(1-\cos
\lambda )\}\sigma _{z}  \label{eq314}
\end{eqnarray}
where we use the Baker-Campbell-Hausdorff formula\cite{Wei}

\begin{eqnarray}
V^{\dagger }(t)\frac{\partial }{\partial t}V(t)=\frac{\partial }{\partial t}%
L+\frac{1}{2!}[\frac{\partial }{\partial t}L,L]+\frac{1}{3!}[[\frac{\partial
}{\partial t}L,L],L]+\frac{1}{4!}[[[\frac{\partial }{\partial t}%
L,L],L],L]+\cdots  \label{eq3141}
\end{eqnarray}
with $V(t)=\exp [L(t)].$ The eigenstates of $\sigma _{z}$ corresponding to
the eigenvalue $\sigma =+1$ and $\sigma =-1$ are ${%
%TCIMACRO{\binom{1 }{0}}%
%BeginExpansion
{1  \choose 0}%
%EndExpansion
}$ and ${%
%TCIMACRO{\binom{0 }{1}}%
%BeginExpansion
{0  \choose 1}%
%EndExpansion
},$ and the eigenstate of $N^{^{\prime }}$ is $%
%TCIMACRO{\binom{\left| m\right\rangle }{\left| m+k\right\rangle}}%
%BeginExpansion
{\left| m\right\rangle  \choose \left| m+k\right\rangle}%
%EndExpansion
$ in terms of (\ref{eq35}). From Eq. (\ref{eq26}), (\ref{eq211}), (\ref
{eq212}), we obtain two particular solutions of the time-dependent
Schr\"{o}dinger equation of the time-dependent TLMJ-C model which are
written in the forms

\begin{equation}
\left| \Psi _{m,\sigma =+1}(t)\right\rangle =\exp \{\frac{1}{i}\int_{0}^{t}[%
\dot{\phi}_{d,\sigma =+1}(t^{^{\prime }})+\dot{\phi}_{g,\sigma
=+1}(t^{^{\prime }})]{\rm d}t^{^{\prime }}\}V(t){%
%TCIMACRO{\binom{\left| m\right\rangle }{0}}%
%BeginExpansion
{\left| m\right\rangle  \choose 0}%
%EndExpansion
}  \label{eq315}
\end{equation}
with
\begin{eqnarray}
\dot{\phi}_{d,\sigma =+1}(t^{^{\prime }}) &=&(m+\frac{k}{2})\omega
(t^{^{\prime }})+\frac{1}{2}\lambda _{m}^{\frac{1}{2}}\{g(t^{^{\prime
}})\exp [-i\gamma (t^{^{\prime }})]+g^{\ast }(t^{^{\prime }})\exp [i\gamma
(t^{^{\prime }})]\}\sin \lambda (t^{^{\prime }})  \nonumber \\
&&-\frac{\delta (t^{^{\prime }})}{2}\cos \lambda (t^{^{\prime }})
\end{eqnarray}
and
\begin{equation}
\dot{\phi}_{g,\sigma =+1}(t^{^{\prime }})=\frac{\dot{\gamma}(t^{^{\prime }})%
}{2}[1-\cos \lambda (t^{^{\prime }})];  \label{eq317}
\end{equation}
and

\begin{equation}
\left| \Psi _{m,\sigma =-1}(t)\right\rangle =\exp \{\frac{1}{i}\int_{0}^{t}[%
\dot{\phi}_{d,\sigma =-1}(t^{^{\prime }})+\dot{\phi}_{g,\sigma
=-1}(t^{^{\prime }})]{\rm d}t^{^{\prime }}\}V(t){%
%TCIMACRO{\binom{0}{\left| m+k\right\rangle }}%
%BeginExpansion
{0 \choose \left| m+k\right\rangle }%
%EndExpansion
}  \label{eq318}
\end{equation}
with
\begin{eqnarray}
\dot{\phi}_{d,\sigma =-1}(t^{^{\prime }}) &=&(m+\frac{k}{2})\omega
(t^{^{\prime }})+\frac{1}{2}\lambda _{m}^{\frac{1}{2}}\{g(t^{^{\prime
}})\exp [-i\gamma (t^{^{\prime }})]+g^{\ast }(t^{^{\prime }})\exp [i\gamma
(t^{^{\prime }})]\}\sin \lambda (t^{^{\prime }})  \nonumber \\
&&+\frac{\delta (t^{^{\prime }})}{2}\cos \lambda (t^{^{\prime }})
\label{eq319}
\end{eqnarray}
and

\begin{equation}
\dot{\phi}_{g,\sigma =-1}(t^{^{\prime }})=-\frac{\dot{\gamma}(t^{^{\prime }})%
}{2}[1-\cos \lambda (t^{^{\prime }})].  \label{eq320}
\end{equation}

These two particular solutions of the Schr\"{o}dinger equation (\ref{eq24})
contain corresponding dynamical phase factor $\exp [\frac{1}{i}\int_{0}^{t}%
\dot{\phi}_{d,\sigma }(t^{^{\prime }}){\rm d}t^{^{\prime }}]$ and
the geometric phase factor $\exp
[\frac{1}{i}\int_{0}^{t}\dot{\phi}_{g,\sigma }(t^{^{\prime
}})]{\rm d}t^{^{\prime }}$ with $\sigma =\pm 1.$ It should be
noted that when the parameter $\lambda $ is taken to be a constant
in the expression (\ref{eq317}) and (\ref{eq320}), the geometric
phases in a cycle in the parameter space of the invariant $I(t)$
can be rewritten as

\begin{equation}
\phi _{\sigma }(T)=\frac{\sigma }{2}2\pi (1-\cos \lambda ),  \label{eq321}
\end{equation}
where $2\pi (1-\cos \lambda )$ is the solid angle over the parameter space
of the invariant $I(t)$ which represents the geometric meanings of the phase
factor. Apparently, it can be seen that the former is dependent on the
transition frequency $\omega _{0}(t)$ and the mode frequency $\omega (t)$,
and the coupling coefficients $g(t)$ and $g^{\ast }(t)$ as well, whereas the
latter is immediately independent of these frequency parameters and the
coupling coefficients.

\section{THE SECOND-QUANTIZED SPIN MODEL AND THE SOLUTION TO THE
TIME-DEPENDENT SCHR\"{O}DINGER EQUATION}

The exact solutions of the time-dependent spin model can be obtained by
using the invariant-related unitary transformation method. The Hamiltonian
of the spin model is generally of the form

\begin{equation}
H(t)=\vec{c}(t)\cdot \vec{J},  \label{eq41}
\end{equation}
where $\vec{c}(t)$ is the time-dependent arbitrary vector parameters and can
be taken
\begin{equation}
\vec{c}(t)=c_{0}(t)[\sin \theta (t)\cos \varphi (t),\sin \theta (t)\sin
\varphi (t),\cos \theta (t)].
\end{equation}
With $J_{\pm }=J_{1}\pm iJ_{2}$ satisfies the commuting relations $%
[J_{3},J_{\pm }]=\pm J_{\pm },[J_{+},J_{-}]=2J_{3}.$ Then the expression (%
\ref{eq41}) for $H(t)$ can be rewritten

\begin{equation}
H(t)=c_{0}(t)\{\frac{1}{2}\sin \theta (t)\exp [-i\varphi (t)]J_{+}+\frac{1}{2%
}\sin \theta (t)\exp [i\varphi (t)]J_{-}+\cos \theta (t)J_{3}\}.
\end{equation}

Note that the analytic solution to the time-dependent Schr\"{o}dinger
equation governing the spin model is easily obtained from the previous
solutions to the time-dependent supersymmetric two-level multiphoton
Jaynes-Cummings. It follows from (\ref{eq31}) and (\ref{eq33}) that if when $%
\omega (t)$ and $N^{^{\prime }}$ are chosen to be $0$ and $1$, respectively,
and the supersymmetric operators, $Q^{\dagger }$, $Q$, and $\frac{1}{2}%
\sigma _{z}$, are respectively replaced with $J_{+},$ $J_{-}$ and $J_{3},$
then the above supersymmetric Jaynes-Cummings is formally changed into the
spin model, and it is therefore convenient to solve this time-dependent spin
model. In other word, we can introduce an invariant-related unitary
transformation operator $V(t)$ as follows

\begin{equation}
V(t)=\exp [\beta (t)J_{+}-\beta ^{\ast }(t)J_{-}]  \label{eq44}
\end{equation}
where the time-dependent parameters are taken to be

\begin{equation}
\beta (t)=-\frac{\lambda (t)}{2}\exp [-i\gamma (t)],\quad \beta ^{\ast }(t)=-%
\frac{\lambda (t)}{2}\exp [i\gamma (t)].
\end{equation}
By making use of the Baker-Campbell-Hausdorff formula\cite{Wei} and $V(t)$
in expression (\ref{eq44}), or immediately from Eq. (\ref{eq314}), one can
obtain $H_{V}(t)$ from $H(t)$

\begin{eqnarray}
H_{V}(t) &=&V^{\dagger }(t)H(t)V(t)-V^{\dagger }(t)i\frac{\partial V(t)}{%
\partial t}  \nonumber \\
&=&\{c_{0}[\cos \lambda \cos \theta +\sin \lambda \sin \theta \cos (\gamma
-\varphi )]+\dot{\gamma}(1-\cos \lambda )\}J_{3}.  \label{eq46}
\end{eqnarray}
Analogous to Eq. (\ref{eq311}), the invariant
\begin{equation}
I(t)=\frac{1}{2}\sin \lambda (t)\exp [-i\gamma (t)]J_{+}+\frac{1}{2}\sin
\lambda (t)\exp [i\gamma (t)]J_{-}+\cos \lambda (t)J_{3}  \label{eq460}
\end{equation}
is changed into the time-independent operator, $J_{3}$, {\it i.e.},
\begin{equation}
I_{V}(t)=V^{\dagger }(t)I(t)V(t)=J_{3},  \label{eq461}
\end{equation}
under this unitary transformation $V(t)$. From the two expressions (\ref
{eq46}) and (\ref{eq461}), one can see that $H_{V}(t)$ differs from $I_{V}$
only by a time-dependent $c$-number factor. Thus with the help of (\ref
{eq210}), (\ref{eq211}) and (\ref{eq212}), it is easy to get the general
solution of the time-dependent Schr\"{o}dinger equation (\ref{eq24})

\begin{equation}
\left| \Psi (t)\right\rangle _{s}=%
%TCIMACRO{\tsum}%
%BeginExpansion
\mathop{\textstyle\sum}%
%EndExpansion
_{m}C_{m}\exp [\frac{1}{i}\phi _{m}(t)]V(t)\left| m\right\rangle
\label{eq47}
\end{equation}
with the coefficients $C_{m}=\langle m,t=0\left| \Psi (0)\right\rangle _{s}.$
The phase $\phi _{m}(t)=\phi _{d,m}(t)+\phi _{g,m}(t)$ includes the
dynamical phase

\begin{eqnarray}
\phi _{d,m}(t) &=&%
%TCIMACRO{\tint}%
%BeginExpansion
\textstyle\int%
%EndExpansion
_{0}^{t}\left\langle m\right| V^{\dagger }(t^{^{\prime }})H(t^{^{\prime
}})V(t^{^{\prime }})\left| m\right\rangle {\rm d}t^{^{\prime }}  \nonumber \\
&=&m%
%TCIMACRO{\tint}%
%BeginExpansion
\textstyle\int%
%EndExpansion
_{0}^{t}c_{0}(t^{^{\prime }})\{\cos \lambda (t^{^{\prime }})\cos \theta
(t^{^{\prime }})+  \nonumber \\
&&+\sin \lambda (t^{^{\prime }})\sin \theta (t^{^{\prime }})\cos
[\gamma (t^{^{\prime }})-\varphi (t^{^{\prime }})]\}{\rm
d}t^{^{\prime }}  \label{eq48}
\end{eqnarray}
and the geometric phase

\begin{equation}
\phi _{g,m}(t)=\int_{0}^{t}\left\langle m\right| -V^{\dagger }(t^{^{\prime
}})i\frac{\partial V(t^{^{\prime }})}{\partial t^{^{\prime }}}\left|
m\right\rangle {\rm d}t^{^{\prime }}=m%
%TCIMACRO{\tint}%
%BeginExpansion
\textstyle\int%
%EndExpansion
_{0}^{t}\dot{\gamma}(t^{^{\prime }})[1-\cos \lambda (t^{^{\prime
}})]{\rm d}t^{^{\prime }}.  \label{eq49}
\end{equation}
\\ \\
In what follows, when the time-dependent arbitrary vector parameters of the
Hamiltonian is chosen to be a special form, the time-dependent
second-quantized spin model is reduced to the generalized Chiao-Wu model
that describes the photon propagating inside the fiber. Consider a
noncoplanarly curved optical fiber that is wound smoothly on a large enough
diameter\cite{Tomita}, a photon propagating inside the fiber is along it at
each point at arbitrary time, one can draw a conclusion that the eigenvalue
of the helicity $\frac{\vec{k}(t)}{k}\cdot \vec{J}$ of the photon is
conserved in motion and its helicity operator $\frac{\vec{k}(t)}{k}\cdot
\vec{J}$ is an invariant. In terms of the invariant equation (\ref{eq21}),
we construct an effective Hamiltonian $H_{eff}(t)$ which represents the
interaction between the photon field and medium of the fiber. The effective
Hamiltonian $H_{eff}(t)$ is of the form\footnote{As far as we are
concerned, the photon propagation problem can be ascribed to a
time-dependent second-quantized spin model, where the effective
(phenomenological) Hamiltonian is of a second-quantized form. Whereas in the
previous
researches\cite{Chiao,Berry,Tomita,Kwiat,Haldane1,Robinson,Haldane2},
this problem was treated often by using classical Maxwell's
Equations and first-quantized Schr\"{o}dinger equation (and
Berry's adiabatic geometric phase formula as well\cite{Berry}).
Although these investigations can be said to be somewhat
outstandingly successful in both predicting and studying adiabatic
geometric phases of photons in the fibre, here I still want to
emphasize two points: for one thing, only by using the
second-quantization formulation can we investigate the photon
geometric phases at quantum level; for another, only when we
consider the non-normal-product second-quantized Hamiltonian can
it enable us to predict the existence of geometric phases at
quantum-vacuum level. Tomita and Chiao may also agree to the above
first point. They held the arguments\cite {Haldane2} that although
the geometric phases in the curved fibre can also be obtained by
means of classical Maxwell's electrodynamics, they preferred to
think of this phenomenon as originating at the quantum level, but
surviving the correspondence-principle limit into the classical
level. However, this point is not the main subject in the present
paper, which will be further discussed elsewhere. Here, instead,
we concentrate only on the second point, {\it i.e.}, the geometric
phases at quantum-vacuum level resulting from the zero-point
radiation fields of vacuum, which has not been investigated in
previous researches.}

\begin{equation}
H_{eff}(t)=\frac{\vec{k}(t)\times \stackrel{\cdot }{\vec{k}}(t)}{k^{2}}\cdot
\vec{J},  \label{eq410}
\end{equation}
where $\vec{J}=\vec{L}+\vec{S}$ is the total angular momentum operator of
second-quantized electromagnetic field, which is given by

\begin{eqnarray}
\vec{L} &=&-\frac{i}{2}%
%TCIMACRO{\tsum}%
%BeginExpansion
\mathop{\textstyle\sum}%
%EndExpansion
_{\lambda }%
%TCIMACRO{\tint }%
%BeginExpansion
\textstyle\int%
%EndExpansion
d\vec{k}\{[(\vec{k}\times \nabla _{\vec{k}})a^{\dagger }(k,\lambda
)]a(k,\lambda )-[(\vec{k}\times \nabla _{\vec{k}})a(k,\lambda )]a^{\dagger
}(k,\lambda )\},  \nonumber \\
\vec{S} &=&-\frac{i}{2}%
%TCIMACRO{\tint }%
%BeginExpansion
\textstyle\int%
%EndExpansion
d\vec{k}\frac{\vec{k}}{k}[a^{\dagger }(k,1)a(k,2)-a^{\dagger
}(k,2)a(k,1)+a(k,2)a^{\dagger }(k,1)-a(k,1)a^{\dagger }(k,2)].  \label{eq411}
\end{eqnarray}
The infinitesimal rotation operator of motion of the photon in the fiber is
given as follows

\begin{equation}
U_{R}=1-i\vec{\vartheta}\cdot \vec{J}  \label{eq412}
\end{equation}
with

\begin{equation}
\vec{\vartheta}=\frac{\vec{k}(t)\times \stackrel{\cdot }{\vec{k}}(t)}{k^{2}}%
{\rm d}t,  \label{eq413}
\end{equation}
where dot denotes the time rate of change of $\vec{k}(t).$ It follows from
Eq. (\ref{eq412}) that the photon state $\left| \sigma ,\vec{k}%
(t)\right\rangle $ ( $\sigma $ denotes the eigenvalue of helicity )
satisfies the following time-evolution equation

\begin{equation}
i\frac{\partial \left| \sigma ,\vec{k}(t)\right\rangle }{\partial t}=\frac{%
\vec{k}(t)\times \stackrel{\cdot }{\vec{k}}(t)}{k^{2}}\cdot \vec{J}\left|
\sigma ,\vec{k}(t)\right\rangle
\end{equation}
with $H_{eff}(t)=\frac{\vec{k}(t)\times \stackrel{\cdot }{\vec{k}}(t)}{k^{2}}%
\cdot \vec{J}$ being the effective Hamiltonian of photon inside the fiber.
Apparently, together with the invariant $I(t)$, the effective Hamiltonian $%
H_{eff}(t)$ agrees with the invariant equation (\ref{eq21}). From the form
of the effective Hamiltonian $H_{eff}(t)$ and the expression (\ref{eq411})
for $\vec{J}$, one can see that the problem of the rotation of the
polarization plane is actually in analogy with that of the time-dependent
second-quantized spin model, and $\frac{\vec{k}(t)\times \stackrel{\cdot }{%
\vec{k}}(t)}{k^{2}}$ may be considered a general magnetic field.

Set the components of momentum of a photon

\begin{equation}
\frac{\vec{k}(t)}{k}=(\sin \lambda (t)\cos \gamma (t),\sin \lambda (t)\sin
\gamma (t),\cos \lambda (t))  \label{eq415}
\end{equation}
where the time-dependent parameters $\lambda (t)$ and $\gamma (t)$ denote
the angle displacement of $\vec{k}(t)$ in the spherical polar coordinate
system. In terms of the above process for getting the solution of the spin
model, we use the invariant-related unitary transformation operator $%
V(t)=\exp [\beta (t)J_{+}-\beta ^{\ast }(t)J_{-}],$ which enables one to
transform the time-dependent invariant $I(t)$ into $I_{V}(t)$ $=J_{3}.$
Under the unitary transformation $V(t)$, one can obtain $H_{V}(t)$ which is
written as (\ref{eq46}) from $H(t)$ by making use of the
Baker-Campbell-Hausdorff formula, where the time-dependent parameters $%
\theta $ and $\varphi $ represent the angle displacement of $\frac{\vec{k}%
(t)\times \stackrel{\cdot }{\vec{k}}(t)}{k^{2}}$ in the spherical polar
coordinate, namely,

\begin{equation}
\frac{\vec{k}(t)\times \stackrel{\cdot }{\vec{k}}(t)}{k^{2}}=(\sin \theta
\cos \varphi ,\sin \theta \sin \varphi ,\cos \theta ).  \label{eq416}
\end{equation}
By using Eq. (\ref{eq415}), (\ref{eq416}) and the invariant equation (\ref
{eq21}), two auxiliary equations can be derived

\begin{equation}
\dot{\gamma}\sin ^{2}\lambda =\cos \theta ,\quad \dot{\lambda}\cos \gamma -%
\dot{\gamma}\cos \lambda \sin \lambda \sin \gamma =\sin \theta \sin \varphi .
\label{eq417}
\end{equation}
From Eqs. (\ref{eq417}), it is shown that in Eq. (\ref{eq46})

\begin{equation}
\cos \lambda \cos \theta +\sin \lambda \sin \theta \cos (\gamma -\varphi )=0,
\label{eq418}
\end{equation}
thus, the expression (\ref{eq46}) can be rewritten as

\begin{equation}
H_{V}(t)=\dot{\gamma}(t)[1-\cos \lambda (t)]J_{3}.  \label{eq420}
\end{equation}
According to Eq. (\ref{eq49}), the geometric phase of a photon whose
eigenvalue of helicity is $\sigma $ can be expressed by

\begin{equation}
\phi _{\sigma }(t)=\{%
%TCIMACRO{\tint}%
%BeginExpansion
\textstyle\int%
%EndExpansion
_{0}^{t}\dot{\gamma}(t^{^{\prime }})[1-\cos \lambda (t^{^{\prime
}})]{\rm d}t^{^{\prime }}\}\left\langle \sigma \right| J_{3}\left|
\sigma \right\rangle .  \label{eq421}
\end{equation}
When we consider the adiabatic case, where $\lambda $ is time-independent
time, $t$, the geometric phase $\phi _{\sigma }$ in a cycle over the
parameter space of the invariant $I(t)$ may be written
\begin{equation}
\phi _{\sigma }(T)=2\pi (1-\cos \lambda ),
\end{equation}
where $\left\langle \sigma \right| J_{3}\left| \sigma \right\rangle =1$ when
the eigenvalue of helicity of the photon is taken to be $\sigma =1.$ It
should be noted that this result agrees with that derived from the Chiao-Wu
theory\cite{Chiao}.

By the aid of (\ref{eq47}) and (\ref{eq421}), it is easy to get the general
solution of the time-dependent Schr\"{o}dinger equation in the fiber
experiment

\begin{equation}
\left| \Psi (t)\right\rangle _{s}=%
%TCIMACRO{\tsum}%
%BeginExpansion
\mathop{\textstyle\sum}%
%EndExpansion
_{\sigma }C_{\sigma }\exp [\frac{1}{i}\phi _{\sigma }(t)]V(t)\left| \sigma
\right\rangle   \label{eq422}
\end{equation}
with the coefficients $C_{\sigma }=\langle \sigma ,t=0\left| \Psi
(0)\right\rangle _{s}.$

It can be seen from the expression (\ref{eq418}) and (\ref{eq420}) that, in
the noncoplanar optical fiber, the dynamical phase of photon due to the
effective Hamiltonian vanishes, and its geometric phase is expressed by the
expression (\ref{eq421}). Since it is a second-quantized spin model, we do
not take the normal product for the third component of the angular momentum $%
\vec{J}$ of the linear polarized photon.

\section{THE EXPLICIT EXPRESSION FOR THE TIME-EVOLUTION OPERATOR}

One of the applications of the invariant-related unitary transformation
formulation is that it can change the hidden form of the time-evolution
operator $U(t)=P\exp [\frac{1}{i}%
%TCIMACRO{\tint}%
%BeginExpansion
\textstyle\int%
%EndExpansion
_{0}^{t}H_{eff}(t^{^{\prime }}){\rm d}t^{^{\prime }}]$ of wave
functions into an explicit expression, where $P$ denotes the
chronological product operator. Assume that the initial momentum
$\vec{k}(t=0)$ is parallel to the third component of the space
coordinate, that is,

\begin{equation}
k_{1}=0,\quad k_{2}=0,\quad k_{3}=k,  \label{eq51}
\end{equation}
we have accordingly

\begin{equation}
\lambda (0)=0,\quad V(0)=1.  \label{eq52}
\end{equation}
Use is made of the content of Sec.II which shows that

\begin{equation}
U(t)=V(t)\exp (fJ_{3}),  \label{eq53}
\end{equation}
where

\begin{equation}
f=\frac{1}{i}%
%TCIMACRO{\tint}%
%BeginExpansion
\textstyle\int%
%EndExpansion
_{0}^{t}\dot{\gamma}(t^{^{\prime }})[1-\cos \lambda (t^{^{\prime
}})]{\rm d}t^{^{\prime }}.  \label{eq54}
\end{equation}
At $t=0$, the eigenstates of $I_{V}$ corresponding to the eigenvalues $%
\sigma =\pm 1$ are of the forms

\begin{eqnarray}
\left| \sigma =+1,\vec{k}(t=0)\right\rangle &=&\frac{1}{\sqrt{2}}[a^{\dagger
}(k,1)+ia^{\dagger }(k,2)]\left| 0\right\rangle \equiv a_{R}^{\dagger
}(k)\left| 0\right\rangle ,  \nonumber \\
\left| \sigma =-1,\vec{k}(t=0)\right\rangle &=&\frac{1}{\sqrt{2}}[a^{\dagger
}(k,1)-ia^{\dagger }(k,2)]\left| 0\right\rangle \equiv a_{L}^{\dagger
}(k)\left| 0\right\rangle  \label{eq55}
\end{eqnarray}
where $a_{R}^{\dagger }(k)$ and $a_{L}^{\dagger }(k)$ are the creation
operators of the right- and left- rotation single-mode polarized photons,
and $k$ denotes the momentum of photon, and $\left| 0\right\rangle $
represents the initial vacuum state at $t=0$. Then $\left| \sigma ,\vec{k}%
(t)\right\rangle $ at arbitrary time is given

\begin{equation}
\left| \sigma ,\vec{k}(t)\right\rangle =U(t)\left| \sigma ,\vec{k}%
(t=0)\right\rangle =V(t)\exp (fJ_{3})a_{\sigma }^{\dagger }\left|
0\right\rangle .  \label{eq56}
\end{equation}

In the following, we show that $\left| \sigma ,\vec{k}(t)\right\rangle $ is
the eigenstate of the momentum operator
\begin{equation}
p_{\mu }=%
%TCIMACRO{\tsum}%
%BeginExpansion
\mathop{\textstyle\sum}%
%EndExpansion
_{\lambda }%
%TCIMACRO{\tint }%
%BeginExpansion
\textstyle\int%
%EndExpansion
d\vec{k}k_{\mu }a^{\dagger }(k,\lambda )a(k,\lambda ).  \label{eq57}
\end{equation}
of photons field. Prior to this, some useful commuting relations are spread
out as follows:

\begin{eqnarray}
\lbrack ip_{2}+p_{1},J_{+}] &=&[ip_{2}-p_{1},J_{-}]=0,  \nonumber \\
\lbrack ip_{2}-p_{1},J_{+}] &=&[ip_{2}+p_{1},J_{-}]=2p_{3},  \nonumber \\
\lbrack J_{3},p_{\mu }] &=&i(\delta _{1\mu }p_{2}-\delta _{2\mu
}p_{1}),[p_{3},J_{\pm }]=ip_{2}\pm p_{1},  \nonumber \\
\lbrack p_{\mu },J_{\pm }] &=&i\delta _{3\mu }p_{2}-i\delta _{2\mu }p_{3}\pm
i(i\delta _{1\mu }p_{3}-i\delta _{3\mu }p_{1}).  \label{eq58}
\end{eqnarray}
Thus, the following formula can be calculated by using the above expressions

\begin{eqnarray}
p_{\mu }U(t)a_{\sigma }^{\dagger }(k)\left| 0\right\rangle  &=&V(t)[p_{\mu
},\exp (fJ_{3})]a_{\sigma }^{\dagger }(k)\left| 0\right\rangle +V(t)\exp
(fJ_{3})p_{\mu }a_{\sigma }^{\dagger }(k)\left| 0\right\rangle +  \nonumber
\\
&&+[p_{\mu },V(t)]\exp (fJ_{3})a_{\sigma }^{\dagger }(k)\left|
0\right\rangle .  \label{eq59}
\end{eqnarray}
First we compute $[p_{\mu },\exp (fJ_{3})].$ By using the formulae (\ref
{eq58}) one will arrive at

\begin{eqnarray}
\exp (-fJ_{3})p_{\mu }\exp (fJ_{3}) &=&i\frac{e^{f}-e^{-f}}{2}(\delta _{2\mu
}p_{1}-\delta _{1\mu }p_{2})+  \nonumber \\
&&+\frac{e^{f}+e^{-f}-2}{2}(\delta _{1\mu }p_{1}+\delta _{2\mu
}p_{2})+p_{\mu },  \label{510}
\end{eqnarray}
then

\begin{eqnarray}
\lbrack p_{\mu },\exp (fJ_{3})] &=&i\frac{e^{f}-e^{-f}}{2}\exp
(fJ_{3})(\delta _{2\mu }p_{1}-\delta _{1\mu }p_{2})+  \nonumber \\
&&+\frac{e^{f}+e^{-f}-2}{2}\exp (fJ_{3})(\delta _{1\mu }p_{1}+\delta _{2\mu
}p_{2}).  \label{eq511}
\end{eqnarray}
Use is made of the expressions (\ref{eq55}), we get

\[
p_{\mu }=%
%TCIMACRO{\tsum}%
%BeginExpansion
\mathop{\textstyle\sum}%
%EndExpansion
_{\lambda }%
%TCIMACRO{\tint }%
%BeginExpansion
\textstyle\int%
%EndExpansion
d\vec{k}k_{\mu }a^{\dagger }(k,\lambda )a(k,\lambda )=%
%TCIMACRO{\tint }%
%BeginExpansion
\textstyle\int%
%EndExpansion
d\vec{k}k_{\mu }\{a_{R}^{\dagger }(k)a_{R}(k)+a_{L}^{\dagger }(k)a_{L}(k)\},
\]
then $p_{\mu }a_{\sigma }^{\dagger }(k)\left| 0\right\rangle =k_{\mu
}a_{\sigma }^{\dagger }(k)\left| 0\right\rangle .$ In terms of the initial
conditions Eq.(\ref{eq52}), the following eigenvalue equations of momentum
can be obtained

\begin{eqnarray}
p_{1}a_{\sigma }^{\dagger }(k)\left| 0\right\rangle &=&k\sin \lambda (0)\cos
\gamma (0)a_{\sigma }^{\dagger }(k)\left| 0\right\rangle =0,  \nonumber \\
p_{2}a_{\sigma }^{\dagger }(k)\left| 0\right\rangle &=&k\sin \lambda (0)\sin
\gamma (0)a_{\sigma }^{\dagger }(k)\left| 0\right\rangle =0,  \nonumber \\
p_{3}a_{\sigma }^{\dagger }(k)\left| 0\right\rangle &=&k\cos \gamma
(0)a_{\sigma }^{\dagger }(k)\left| 0\right\rangle =ka_{\sigma }^{\dagger
}(k)\left| 0\right\rangle .  \label{eq512}
\end{eqnarray}
By making use of the expressions (\ref{eq511}) and (\ref{eq512}), the first
term on the right-hand side in (\ref{eq59}) is evidently

\begin{equation}
V(t)[p_{\mu },\exp (fJ_{3})]a_{\sigma }^{\dagger }(k)\left| 0\right\rangle =0
\label{eq513}
\end{equation}
and the second term is expressed by

\begin{equation}
V(t)\exp (fJ_{3})p_{\mu }a_{\sigma }^{\dagger }(k)\left| 0\right\rangle
=k_{\mu }V(t)\exp (fJ_{3})a_{\sigma }^{\dagger }(k)\left| 0\right\rangle .
\label{eq514}
\end{equation}
Next we calculate $[p_{\mu },V(t)].$ Using the commuting relations (\ref
{eq58}), the complicated calculation yields

\begin{eqnarray}
V^{\dagger }p_{\mu }V &=&\exp \{-[\beta (t)J_{+}-\beta ^{\ast
}(t)J_{-}]\}p_{\mu }\exp [\beta (t)J_{+}-\beta ^{\ast }(t)J_{-}]  \nonumber
\\
&=&p_{\mu }+\frac{\sin \lambda }{\lambda }(\Delta _{\mu }p_{3}+\delta _{3\mu
}C)+\frac{1}{\lambda ^{2}}(1-\cos \lambda )(\Delta _{\mu }C-4\beta \beta
^{\ast }\delta _{3\mu }p_{3})  \label{eq515}
\end{eqnarray}
where

\begin{eqnarray}
\Delta _{\mu } &=&\beta ^{\ast }(i\delta _{2\mu }-\delta _{1\mu })-\beta
(i\delta _{2\mu }+\delta _{1\mu }),  \nonumber \\
C &=&\beta ^{\ast }(p_{1}-ip_{2})+\beta (p_{1}+ip_{2}).  \label{eq516}
\end{eqnarray}
$[p_{\mu },V(t)]$ can be derived from Eq. (\ref{eq515}). Based on this, one
can derive the third term on the right-hand side in (\ref{eq59}), which may
be rewritten as

\begin{eqnarray}
\lbrack p_{\mu },V]\exp (fJ_{3})a_{\sigma }^{\dagger }(k)\left|
0\right\rangle &=&\frac{\sin \lambda }{\lambda }\Delta _{\mu }V\exp
(fJ_{3})p_{3}a_{\sigma }^{\dagger }(k)\left| 0\right\rangle  \nonumber \\
&&-(1-\cos \lambda )\delta _{3\mu }V\exp (fJ_{3})p_{3}a_{\sigma }^{\dagger
}(k)\left| 0\right\rangle .  \label{eq517}
\end{eqnarray}
According to Eq.(\ref{eq512}) and (\ref{eq516}), we can arrive at

\begin{equation}
p_{\mu }U(t)a_{\sigma }^{\dagger }(k)\left| 0\right\rangle =[k(\delta _{1\mu
}\cos \gamma +\delta _{2\mu }\sin \gamma )\sin \lambda -k(1-\cos \lambda
)\delta _{3\mu }+k_{\mu }]U(t)a_{\sigma }^{\dagger }(k)\left| 0\right\rangle
.  \label{eq518}
\end{equation}

By combining Eq. (\ref{eq518}) with (\ref{eq512}), the eigenvalue equation
of momentum of photons field can be rewritten in the form

\begin{equation}
p_{\mu }U(t)a_{\sigma }^{\dagger }(k)\left| 0\right\rangle =k_{\mu
}(t)U(t)a_{\sigma }^{\dagger }(k)\left| 0\right\rangle  \label{eq519}
\end{equation}
where $k_{1}(t)=k\sin \lambda \cos \gamma ,$ $k_{2}(t)=k\sin \lambda \sin
\gamma ,$ $k_{3}(t)=k\cos \lambda .$ We thus show that $U(t)a_{\sigma
}^{\dagger }(k)\left| 0\right\rangle $ is truly the eigenstate of momentum
operator of the second-quantization photons field, and the eigenvalue is the
time-dependent momentum $k_{\mu }(t)$ of the single-mode photon.

In accordance with the L-R invariant theory, the wave function $\left|
\sigma ,\vec{k}(t)\right\rangle =U(t)a_{\sigma }^{\dagger }(k)\left|
0\right\rangle $ of the single-mode photon is also the eigenstate of the
invariant $I(t)=\frac{\vec{k}(t)}{k}\cdot \vec{J}.$ The significance of the
invariant-related unitary transformation method will be illustrated in the
following, by using the explicit expression for the time-evolution operator.
The applications of the following commuting relations

\begin{eqnarray}
\lbrack J_{1,}V] &=&\sin \lambda \cos \gamma VJ_{3}+\frac{\cos \gamma
(1-\cos \lambda )}{\lambda }V(\beta ^{\ast }J_{-}+\beta J_{+}),  \nonumber \\
\lbrack J_{2,}V] &=&\sin \lambda \sin \gamma VJ_{3}+\frac{\sin \gamma
(1-\cos \lambda )}{\lambda }V(\beta ^{\ast }J_{-}+\beta J_{+}),  \nonumber \\
\lbrack J_{3,}V] &=&(\cos \lambda -1)VJ_{3}+\frac{\sin \lambda }{\lambda }%
V(\beta ^{\ast }J_{-}+\beta J_{+})  \label{eq520}
\end{eqnarray}
enable one to get

\begin{eqnarray}
\lbrack \frac{\vec{k}(t)}{k}\cdot \vec{J},V]\exp (fJ_{3}) &=&U(t)J_{3}-V%
\frac{\vec{k}(t)}{k}\cdot \vec{J}\exp (fJ_{3})  \nonumber \\
&=&U(t)J_{3}-VI(t)\exp (fJ_{3}).  \label{eq521}
\end{eqnarray}
Since

\begin{eqnarray}
\frac{\vec{k}(t)}{k}\cdot \vec{J}U(t)a_{\sigma }^{\dagger }(k)\left|
0\right\rangle &=&V[\frac{\vec{k}(t)}{k}\cdot \vec{J},\exp
(fJ_{3})]a_{\sigma }^{\dagger }(k)\left| 0\right\rangle +  \nonumber \\
&&+[\frac{\vec{k}(t)}{k}\cdot \vec{J},V]\exp (fJ_{3})a_{\sigma }^{\dagger
}(k)\left| 0\right\rangle +  \nonumber \\
&&+U(t)\frac{\vec{k}(t)}{k}\cdot \vec{J}a_{\sigma }^{\dagger }(k)\left|
0\right\rangle ,  \label{eq522}
\end{eqnarray}
we apply the initial conditions (\ref{eq52}) and (\ref{eq512}) to Eq.(\ref
{eq522}) and obtain

\begin{equation}
\frac{\vec{k}(t)}{k}\cdot \vec{J}U(t)a_{\sigma }^{\dagger }(k)\left|
0\right\rangle =\sigma U(t)a_{\sigma }^{\dagger }(k)\left| 0\right\rangle
\label{eq523}
\end{equation}
with the eigenvalue of $\frac{\vec{k}(t)}{k}\cdot \vec{J}$ being $\sigma
=\pm 1$ corresponding to the right- and left- rotation linear polarized
photons, respectively.

From what has been discussed above, one can draw a conclusion that one of
the advantages of the invariant-related unitary transformation method is
transforming the evolution operator $U(t)$ of hidden form to the explicit
expression (\ref{eq53}).

\section{CONCLUDING REMARKS}

We construct an invariant in the sub-Hilbert-space corresponding to a
particular eigenvalue of the time-independent invariant $N^{^{\prime }}$ and
get the exact solutions of the time-dependent TLMJ-C model by making use of
the invariant-related unitary transformation formulation. In view of the
above calculation, we can see that this unitary transformation formulation
has some useful applications, for instance, it can solve the time-dependent
systems and treat the geometric phase factor, and obtain the explicit
expressions, instead of the hidden form, for the evolution operator of the
wave functions.

Since the three-level two-mode Jaynes-Cummings model plays an important role
in Quantum Optics, the supersymmetric structure and the exact solutions of
the time-dependent three-level two-mode multiphoton J-C model deserves
further investigations by the formalism suggested in the present paper.

We construct an effective Hamiltonian in this paper since the helicity of
the photon field is a pseudo scalar and cannot be regarded as the
Hamiltonian. We transform the problem of motion of the photon in the optical
fiber into that of the time-dependent quantum spin model. Effective
Hamiltonian is obtained by making use of the invariant equation (\ref{eq21}%
), rather than through analyzing the electromagnetic interaction between the
photons and the medium of the optical fiber.

The invariant-related unitary transformation formulation is an effective
method for treating the geometric phase factor\cite{Gao2,Gao4}. This
formulation replaces eigenstates of the time-dependent invariants by that of
the time-independent invariants through the unitary transformation. It uses
the invariant-related unitary transformation and obtain the explicit
expression for the time-evolution operator, instead of the formal solution
that is related to the chronological product. 
\\

{\it Note added}: In addition, we also treat the
non-normal-order spin operators and consider the potential effects
({\it e.g.}, {\it quantum-vacuum geometric phases}) of quantum
fluctuation fields arising in a curved optical fibre. The
quantum-vacuum geometric phase, which is of physical interest, can
be deducted by using the operator normal product, and the doubt of
validity and universality for the normal-normal procedure applied
to time-dependent quantum systems is thus proposed. Our brief history of investigating photon geometric phases in the
fibre is as follows: in April 2000, Gao and I began to consider
the {\bf non-cyclic non-adiabatic geometric phases} of photon
fields in the curved fibre based on a {\bf second-quantized spin
model}. In May 2000,
Gao first proposed the concept of {\bf quantum-vacuum geometric
phases}. The existence problem of quantum-vacuum geometric phases
is strongly relevant to whether the second quantization in spin
model is adopted or not. Since it is in connection with properties
of quantum electromagnetic vacuum and, moreover, this geometric
phase is related close to the topological and global features of
time evolution of vacuum-fluctuation fields, we think this concept
is of essential significance and therefore deserves detailed
investigations. From then on, these problems gained our attention
and we tried to investigate this topological phases at
quantum-vacuum level.

Since quantum-vacuum geometric phases has an important connection
with vacuum energies, these experimental realizations may be
relevant to the validity problem of normal-product procedure in
the time-dependent quantum field theory (TDQFT), {\it i.e.}, we
also aim to re-examine the normal-product procedure in some
extensions. If the {\it quantum-vacuum geometric phases} is proved
present experimentally, then it is reasonably believed that it is
not suitable for us to remove vacuum fluctuation energies and
infinite charge density just by using the old formulation, {\it
e.g.}, re-defining the vacuum background energies and electric
charges by utilizing the normal-product procedure, since in this
re-definition, some potential physically interesting vacuum
effects may also be removed theoretically. We think only for the
{\it time-independent} quantum field systems can we use safely the
normal-product procedure without any fear of introducing any new
problems other than those which quantum field theory had
encountered before. However, for the {\it
time-dependent} quantum field systems, ({\it e.g.}, photon fields
propagating inside a helically curved fibre and quantum fields in
an expanding universe), the physically interesting vacuum effects
will unfortunately be deducted by the second-quantization
normal-order formulation. So, the normal-order technique may
therefore not be applicable to the {\it time-dependent} quantum
fields. To the best of our knowledge, in the literature, this
normal order problem in the time-dependent quantum field theory
gets less attention and interests than it deserves. To test our
above theoretical viewpoints, we hope the quantum-vacuum geometric
phases of photons in the curved fibre would be investigated
experimentally in the near future.

Unfortunately, the left-handed polarized light due to vacuum
fluctuation is often accompanied by the zero-point right-handed
polarized light and their total quantum-vacuum geometric phases is
therefore vanishing. So, it is not easy for physicists to
investigate experimentally the quantum-vacuum geometric phases.
This, therefore, means that our above theoretical remarks as to
whether the normal-product procedure is valid or not for the
time-dependent quantum field theory (TDQFT) cannot be examined
experimentally. During the last three years, I tried my best but
unfortunately failed to suggest an excellent idea of experimental
realization of quantum-vacuum geometric phases. We conclude that
it seems not quite satisfactory to test the quantum-vacuum
geometric phases by using the optical fibre that is made of
isotropic media, inhomogeneous media ({\it e.g.}, photonic
crystals\footnote{Photonic crystals are artificial materials
patterned with a periodicity in dielectric constant, which can
create a range of forbidden frequencies called a photonic band
gap. Such dielectric structure of crystals offers the possibility
of molding the flow of light (including the zero-point
electromagnetic fields of vacuum). It is believed that, in the
similar fashion, this effect ({\it i.e.}, modifying the mode
structures of vacuum electromagnetic fields) may also take place
in gyrotropic media. This point will be applied to the discussion
which follows.}), left-handed media (a kind of artificial
composite metamaterial with negative refractive index), uniaxial
(biaxial) crystals or chiral materials. Is it truly extremely
difficult to realize such a goal? It is found finally that perhaps
in the fibre composed of some anisotropic media (such as
gyrotropic materials, including also gyroelectric and gyromagnetic
media, where both electric permittivity and magnetic permeability
are respectively the tensors) the quantum-vacuum geometric phases
may be achieved test experimentally. In these gyrotropic media,
only one of the LRH polarized lights can be propagated without
being absorbed by media. This result holds also for the zero-point
electromagnetic fields. It is well known that people can
manipulate vacuum so as to alter the zero-point mode structures of
vacuum, which has been illustrated in photonic crystals and
Casimir's effect (additionally, the space between two parallel
mirrors, cavity in cavity QED, {\it etc.}). If, for example, in
some certain gyrotropic media one of the LRH polarized lights,
say, the left-handed polarized light, dissipates due to the medium
absorption and only the right-handed light is allowed to be
propagated (in the meanwhile the mode structure of vacuum in these
anisotropic media also alters correspondingly), then the
quantum-vacuum geometric phase of right-handed polarized light can
be easily tested in the fibre fabricated from these gyrotropic
media. 
\\ \\

In order to
illustrate our brief history (April, 2000 $\sim $ March, 2003) of investigating geometric phases of photons in the noncoplanarly curved fibre, an outline is given
as follows:
\\ \\

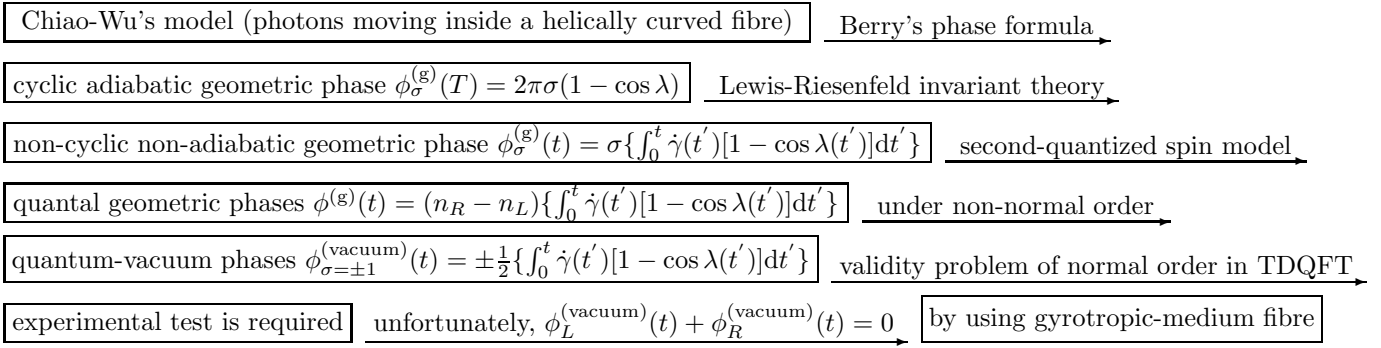
 \setlength{\unitlength}{1cm}
\begin{picture}(30,5)
\put(0,4){\framebox(10.7,0.5){Chiao-Wu's model (photons moving
inside a helically curved fibre)}} \put(10.9,4){\vector(1,0){3.8}}
\put(11.1,4.1){Berry's phase formula}
\put(0,3.2){\framebox(9.1,0.5){cyclic adiabatic geometric phase
$\phi _{\sigma }^{\rm (g)}(T)=2\pi\sigma(1-\cos \lambda )$}}
\put(9.3,3.2){\vector(1,0){5.5}} \put(9.5,3.3){Lewis-Riesenfeld
invariant theory} \put(0,2.4){\framebox(12.3,0.5){non-cyclic
non-adiabatic geometric phase $\phi _{\sigma }^{\rm
(g)}(t)=\sigma\{{\int_{0}^{t}\dot{\gamma}(t^{^{\prime }})[1-\cos
\lambda (t^{^{\prime }})]{\rm d}t^{^{\prime }}}\}$}}
\put(12.5,2.4){\vector(1,0){4.8}} \put(12.7,2.5){second-quantized
spin model}

\put(0,1.6){\framebox(11.2,0.5){quantal geometric phases
$\phi^{\rm
(g)}(t)=(n_{R}-n_{L})\{{\int_{0}^{t}\dot{\gamma}(t^{^{\prime
}})[1-\cos \lambda (t^{^{\prime }})]{\rm d}t^{^{\prime }}}\}$}}
\put(11.4,1.6){\vector(1,0){4.1}} \put(11.6,1.7){under non-normal
order}

\put(0,0.8){\framebox(10.8,0.6){quantum-vacuum phases
$\phi_{\sigma=\pm 1}^{\rm
(vacuum)}(t)=\pm\frac{1}{2}\{{\int_{0}^{t}\dot{\gamma}(t^{^{\prime
}})[1-\cos \lambda (t^{^{\prime }})]{\rm d}t^{^{\prime }}}\}$}}
\put(11,0.8){\vector(1,0){7.1}} \put(11.1,0.9){validity problem of
normal order in TDQFT}

\put(0,0){\framebox(4.6,0.5){experimental test is required}}
\put(4.8,0){\vector(1,0){7.2}} \put(4.9,0.15){unfortunately,
$\phi_{L}^{\rm (vacuum)}(t)+\phi_{R}^{\rm (vacuum)}(t)=0$
}\put(12.2,0){\framebox(5.3,0.6){by using gyrotropic-medium
fibre}}
\end{picture}                                                                                     
\\ \\

{\bf Acknowledgments} The authors thank Prof. X.C. Gao for the helpful discussion.
This project was supported by the National Natural Science Foundation of
China under the project No. $30000034$.

\end{document}